\documentstyle[prl,aps,epsfig]{revtex}

\newcommand \be{\begin{equation}}
\newcommand \ee{\end{equation}}
\newcommand \beq{\begin{eqnarray}}
\newcommand \eeq{\end{eqnarray}}
\newcommand{\set}[2]{\newcommand{#1}{#2}}
\set{\pa}{\partial \over \partial\, }
\set{\leftvector}{\stackrel{\leftarrow}{\partial }}
\set{\rightvector}{\stackrel{\rightarrow}{\partial }}

\begin{document}
\twocolumn[\hsize\textwidth\columnwidth\hsize
           \csname @twocolumnfalse\endcsname

\title{TOPOLOGY OF EVENT DISTRIBUTIONS AS A GENERALIZED DEFINITION 
OF PHASE TRANSITIONS IN FINITE SYSTEMS}
\author{Ph. CHOMAZ$^{(1)}$, F. GULMINELLI$^{(2)}$ and V. DUFLOT$^{(1,2)}$}
\address{(1) GANIL (DSM-CEA/IN2P3-CNRS), B.P.5027, F-14021 Caen cedex,
France \\
 (2) LPC Caen, (IN2P3-CNRS/ISMRA et Universit\'e), F-14050 Caen cedex, France}
\maketitle

\begin{abstract}
We  propose a definition of phase transitions in 
finite systems based on topology anomalies 
of the event distribution in the space of observations. 
This generalizes all the definitions based on the curvature 
anomalies of thermodynamical potentials and  provides a
natural definition of order parameters.
It is directly operational 
from the experimental point of view. It allows to study phase transitions in
Gibbs equilibria as well as in 
other ensembles such as the Tsallis ensemble. 
\end{abstract}

\vskip2pc
]



Phase transitions are amazing examples of self organization of nature. Their
universal character is patent. They are observed at all scales. Elementary
particles deconfine in particle accelerators. Water boils in kettles. Self
gravitating systems collapse in the cosmos. 

From the theoretical point of view phase transitions are defined on very
robust foundations in the thermodynamical limit. 
However, many (and maybe most) physical situations fall out of this
theoretical framework because the thermodynamical limit conditions cannot be
fulfilled. The forces might not be saturating such as the gravitational or
the Coulomb forces. The system might be too small such as any mesoscopic
system. The statistical ensemble might not be of Boltzmann-Gibbs type such
as in Tsallis ensembles, or in non ergodic (or non mixing) systems or even
in collections of events prepared in a dynamical way. In all these cases, a
proper definition and study of phase transitions far from the
thermodynamical limit should be achieved. 

The astrophysics community discussed 
the existence of a microcanonical negative specific heat for collapsing
self-gravitating systems\cite{gravity}. This idea was then extended to the
melting and the boiling of clusters \cite{labastie}. 
In the nuclear multifragmentation context \cite{SMM_MMMC} the phase
transition has first been related to anomalies in the caloric curve. 
The first experimental evidences for such a negative heat capacity have been
reported in the last months \cite{michela-c,INDRA,haberland}. This idea has
been generalized to any statistical ensemble with at least one extensive
variable allowing to sample the coexistence region\cite{Bormio}.

In this paper, we propose a general definition of phase
transitions based on anomalies of the probability distribution of observable
quantities. 
From the theory side, this allows to extend the already given definition\cite
{Bormio} to any situations even out of Boltzmann-Gibbs equilibria. It clarifies
the respective role of order and control parameters. From the experimental
point of view, this new definition gives a way to identify the order
parameter and to extract the meaningful thermodynamical potential and
equation of states.

The order parameter is a quantity which can be known for every single event 
$(n)$
of the considered statistical ensemble, $\xi =\left\{ n\right\} .$
It is an observable which clearly separates the two phases. It is not
necessarily unique. Let us consider a set of $K$ independent observables, $%
\hat{B}_{{\sl k}},$ which form a space containing one possible order
parameter. We can sort events according to the results of the measurement $%
{\bf b}^{\left( n\right) }{\bf \equiv }\left( b_{k}^{\left( n\right)
}\right) $ and thus define a probability distribution of the observables $%
P_{\xi }\left( {\bf b}\right) .$ %

Within the quantum mechanics framework, the statistical ensemble $\xi $ is
described by the density matrix 
$\hat{D}_{\xi }\equiv \sum_n \left| \Psi _{\xi
}^{\left( n\right) }\right\rangle \;p_{\xi }^{\left( n\right)
}\;\left\langle \Psi _{\xi }^{\left( n\right) }\right| $. The states$\left|
\Psi _{\xi }^{\left( n\right) }\right\rangle $ are elements of the Fock
subspace, ${\cal F},$ of the system.%
The observables $\hat{B}_{{\sl k}}$ are operators defined on ${\cal F}$. 
For simplicity we will assume that all the observed operators commute. 
The result of a measurement on the $n-th$ event is $b_{{\sl k}}^{\left(
n\right) }=\left\langle \Psi ^{\left( n\right) }\right| \hat{B}_{{\sl k}%
}\left| \Psi ^{\left( n\right) }\right\rangle $, and so the probability
distribution of the results of the observation ${\bf b}$ reads 
\begin{equation}
p_{\xi }\left( {\bf b}\right) ={\rm Tr}\hat{D}_{\xi }\delta \left( {\bf b-%
\hat{B}}\right) \equiv <\delta \left( {\bf b-\hat{B}}\right) > .
\end{equation}
%
Typical examples of order parameters 
are one body operators such as the density (or
the mean radius) for the liquid gas phase transition or the magnetization in
the ferromagnetic transition. One may also use the dynamical response of the
system to an external excitation $\hat{F}\exp (-i\omega t)+h.c.$ which 
corresponds to the observable $\hat{B}\left( \omega \right) =$ $\hat{F}%
\delta \left( {\sl \hat{H}}-\omega \right) \hat{F}.$ The response of the
system can also be characterized by its moments associated with 
the average value of $\hat{B}_{k}=$ $\left[ \hat{F}\left[ \hat{H}^{k},\hat{F}%
\right] \right] .$

We propose to define phase transitions through the topology of the
probability distribution $P_{\xi }\left( {\bf b}\right) .$ In the absence of
a phase transition $P_{\xi }\left( {\bf b}\right) $ is expected to be normal
and $\log P_{\xi }\left( {\bf b}\right) $ concave. Any abnormal (e.g. bimodal)
behavior of $%
P_{\xi }\left( {\bf b}\right) $ or any convexity anomaly of $\log P_{\xi
}\left( {\bf b}\right) $ signals a phase transition. 
More specifically, the larger eigenvalue of the tensor 
\begin{equation}
T_{\xi }^{k,k^{\prime }}\equiv \frac{\partial ^{2}\log P_{\xi }\left( {\bf b}%
\right) }{\partial b_{k}\partial b_{k^{\prime }}}  \label{Eq:tensor}
\end{equation}
becomes positive in presence of a first order phase transition\cite{gross}. The
associated eigenvector defines the local order parameter since it allows the
best separation of the probability $P_{\xi }\left( {\bf b}\right) $ into two
components which can be recognized as the precursors of phases which will
appear in the thermodynamical limit. 
If the largest eigenvalue is zero, the number of higher derivatives which
are also zero defines the order of the phase transition.

For 
a unique observable $\hat{B},$ the above
definition tells us that when the probability is bimodal we are in presence
of a phase coexistence. The observable $\hat{B}$ is then the order
parameter. In a multidimensional space if the ensemble of events splits into
two components then we are also in presence of a (first order) phase
coexistence. The axis allowing to make a best separation of the event cloud
into two components is an order parameter. Many tools such as the principal
component analysis already exist to perform this topological analysis of the
event distribution\cite{Desesquelle}. 

The definition of phase transition from the
topology of $p_{\xi }\left( {\bf b}\right) $ contains and generalizes all
the definitions based on convexity anomalies of thermodynamical potentials.
Any Boltzmann-Gibbs equilibrium is obtained by maximizing the Shannon
information entropy $S\equiv {\rm Tr}\hat{D}\log \hat{D}$ in the given Fock
space ${\cal F}$ under the constraints of the various observables $\hat{B}_{%
{\sl k}}$ known in average. A Lagrange multiplier $\alpha _{k}$ is
associated with every constraint.\ We assume that the observables are either
known in average $<\hat{B}_{{\sl k}}>=b_{{\sl k}}$ 
or not constrained ($\alpha _{k}=0).$ Other constraints can be applied to
the system through conservation laws on the accessible space ${\cal F}$ 
or through additional Lagrange multipliers $\lambda _{\ell }$ if some other
observable $\hat{A}_{\ell }$ (not related to the order parameter) has an
expectation value known in average 
or imposed by a reservoir. The statistical ensemble is thus defined as $\xi
\equiv \left( {\cal F},{\bf \lambda ,\alpha }\right) $ and its density
matrix reads
\begin{equation}
\hat{D}_{{\cal F}{\bf \lambda \alpha }}=\frac{1}{Z_{%
{\cal F}{\bf \lambda \alpha }}}\exp \left( -\sum_{\ell =1}^{L}\lambda _{\ell
}\hat{A}_{\ell }-\sum_{k=1}^{K}\alpha _{k}\hat{B}_{k}\right) .
\label{Eq:D}
\end{equation}
This ensemble is consistent with the fact that the order parameter is in
general not controlled on an event by event basis but measured. It
spontaneously takes a non zero average value in one (or both) of the two
phases.

It is easy to demonstrate that 
$P_{\xi }\left( {\bf b}\right) $ can be written as 
\begin{equation}
\log P_{{\cal F}{\bf \lambda \alpha }}\left( {\bf b}\right) =\log \bar{W}_{%
{\cal F}{\bf \lambda }}\left( {\bf b}\right) -\sum_{k=1}^{K}\alpha
_{k}b_{k}-\log Z_{{\cal F}{\bf \lambda \alpha }}  
\label{Eq:proba}
\end{equation}
where $\bar{W}_{{\cal F}{\bf \lambda }}\left( {\bf b}\right) =Z_{{\cal F}%
{\bf \lambda 0}}P_{{\cal F}{\bf \lambda 0}}\left( {\bf b}\right) $ 
is nothing but the partition sum of the statistical ensemble associated with
fixed values, ${\bf b,}$ of all the observables. 
Indeed, 
the two partition sums are related through the usual Laplace transform%
\begin{equation}
Z_{{\cal F}{\bf \lambda \alpha }}=\int d{\bf b}\;\bar{W}_{{\cal F}{\bf %
\lambda }}\left( {\bf b}\right) \exp (-{\bf \alpha b}).
\end{equation}
Eq. (\ref{Eq:proba}) clearly demonstrates that the study of convexity
anomalies 
of $\log P_{{\cal F}{\bf \lambda
\alpha }}\left( {\bf b}\right) $ for any value of the  variables ${\bf %
\alpha }$ is equivalent to the study of the curvature anomalies of the
thermodynamical potential $\log \bar{W}_{{\cal F}{\bf \lambda }}\left( {\bf b%
}\right) $ for which the ${\bf b}$ are the control parameters. 
The equations of state related to the partition sum 
$\bar{W}_{{\cal F}{\bf \lambda }}$
can be obtained from the probability distribution
using Eq. (\ref{Eq:proba}) through   
\begin{equation}
\bar{\alpha}_k\left( {\bf b}\right) \equiv 
\frac{\partial \log \bar{W}_{{\cal F}{\bf \lambda }}
\left( {\bf b}\right) 
}{ \partial b_k}
=\frac{\partial \log P_{{\cal F}{\bf \lambda \alpha }}
\left( {\bf b}\right)}{\partial b_k}+\alpha_{k} .
\label{Eq:alpha_bar}
\end{equation}
It presents a back-bending in the abnormal curvature region. There, one 
$\bar{\alpha}_k$ is associated to three values of $b_k$. This is not
the case for the equation of state of the ensemble  (\ref{Eq:D})
$<b_k>_{{\cal F}{\bf \lambda \alpha }}=
-\partial Z_{{\cal F}{\bf \lambda \alpha }} / \partial \alpha_k$
for which only one $<b_k>$ can be associated to one $\alpha_k$.
Conversely, in the regions where the probability distribution is normal
the average $<{\bf b}>$ is expected to be close 
to 
the most probable ${\bf b}_{max}$  
characterized by 
$\bar{\alpha}_k\left( {\bf b}_{max}\right)=\alpha_{k}$.


\begin{figure}[tbph]
\begin{center}
\epsfig{file=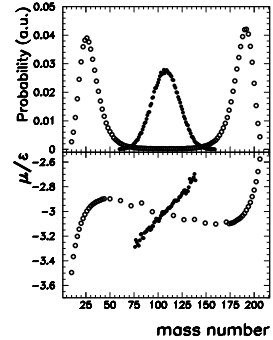,height=6.0cm}
\end{center}
\caption{Grancanonical lattice gas results at $\mu =-3\epsilon $ and $%
T<T_{c}$ (open symbols), $T>T_{c}$ (filled symbols). 
Top: total mass distribution. Bottom:
canonical equation of states (see text).}
\end{figure}

Let us take first the example of the energy as a possible order parameter
with no other constraints, $\hat{B}_{1}{\bf =}\hat{H}$ and $b_{1}=e$ . Then
the considered ensemble is nothing but the canonical one with $\alpha
_{1}=\beta $, the inverse of the temperature. The canonical probability reads 
\begin{equation}
P_{\beta }\left( e\right) =\exp \left( S(e)-\beta e-\log Z\left( \beta
\right) \right) 
\end{equation}
where the entropy, $S(e),$ is related to the level density by $S(e)=\log 
\bar{W}\left( e\right) $. 
A convex intruder in $S\left( e\right) $ directly induces a convexity
anomaly in $\log P_{\beta }\left( e\right) $ which becomes bimodal in the
phase transition region. 
Therefore the definition of phase transition through the curvature anomalies
or a bimodality in the canonical probability distribution 
contains the former definitions based on the occurrence of negative heat
capacities\cite{labastie,gross,haberland,europhys}, the only condition being
that the canonical ensemble exists. 

As a second example we consider 
the grand canonical distribution of
particles. We introduce $\hat{A}_{1}=\hat{H}$ and $\hat{B}_{1}=\hat{N}$ .
Taking $\lambda_1 =\beta $ and $\alpha_1 =-\beta \mu $ we recover the usual
definitions of the temperature and chemical potential. We present 
results from the grand canonical lattice-gas model with fixed volume and
periodic boundary conditions \cite{lee} (see ref. \cite{Bormio} for 
details) with a closest neighbour interaction $-\varepsilon .$ %
In the following the chemical potential will be kept fixed at its critical
value $\mu _{c}=-3\varepsilon $. 
Above the critical temperature the distribution of particle number, $%
P_{\beta \mu }\left( n\right) $ is normal. 
Below the
critical temperature the probability distribution becomes bimodal and
signals the phase transition (see Fig. 1). Indeed 
\begin{equation}
\log P_{\beta \mu }\left( n\right) =\log \bar{Z}_{\beta }\left( n\right)
+\beta \mu n-Z_{\beta \mu }
\end{equation}
where $\bar{Z}_{\beta }\left( n\right) $ is the canonical partition sum for $%
n$ particles while $Z_{\beta \mu }$ is the grand canonical one. The
canonical chemical potential 
is given by 
\begin{equation}
\bar{\mu}_{\beta }\left( n\right) 
\equiv -\beta ^{-1} \frac{\partial \log 
\bar{Z}_{\beta }\left( n\right) }{\partial n}
=-\beta ^{-1}\frac{\partial \log P_{\beta
\mu }\left( n\right) }{\partial n}+\mu 
\end{equation}
and is shown in the lower part of Fig. 1. It should be noticed that a
unique grand canonical chemical potential $\mu $ gives access to the whole
distribution of canonical chemical potentials $\bar{\mu}_{\beta }\left(
n\right) .$ In the phase transition region $\bar{\mu}_{\beta }$ presents a
strong back bending (see fig 1) which comes from the bimodal structure of
the probability distribution and which signals the phase transition. 


\begin{figure}[htbp]
\begin{center}                                     
\epsfig{file=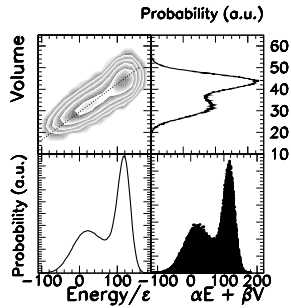,height=6.0cm}
\end{center}
\caption{Volume and energy distribution of a confined canonical lattice-gas
model in the first order phase transition region with three associated
projections. }
\end{figure}

In the previous examples both the energy and the particle number are
conserved quantities. However there is no reason that the order parameter is
associated with any conservation rule. Let us take the example of the
liquid-gas phase transition in an open system of $N$ particles for which
only the average volume is known. In such a case we can define an observable 
$\hat{B}_{1}$ as a measure of the size of the system; for example the cubic
radius 
$\hat{B}_{1}=\frac{4\pi }{3N}\sum_{i}\hat{r}_{i}^{3}\equiv \hat{V}$ where
the sum runs over all the particles $.$ Then a Lagrange multiplier $\lambda
_{V}$ has to be introduced which has the dimension of a pressure divided by
a temperature. 
In a canonical ensemble with an inverse
temperature $\beta$ 
we can define
different distributions which are illustrated in Fig. 2. %
A complete information is contained in the distribution $P_{\beta
\lambda _{v}}\left( e,v\right) =\bar{W}\left( e,v\right) Z_{\beta \lambda
_{v}}^{-1}\exp -(\beta e+\lambda _{v}v)$ since events are sorted according
to the two thermodynamical variables, $e$ and $v$. 
This leads to the density of states $%
\bar{W}\left( e,v\right) $ with a volume $v$ and an energy $e.$ One can see
that in the first order phase transition region the probability distribution
is bimodal. 
In principle one could use the tensor (\ref{Eq:tensor}) to define the
topology so that the order parameter axis 
corresponds to the ridge passing through the saddle point between the liquid
and the gas peaks. In the spirit of the principal component analysis we can
look for an order parameter $\hat{Q}=x\hat{H}+y\hat{V}$ which provides the best
separation of the two phases. A projection of the event on this order
parameter axis is also shown in Fig. 2. One can see a clear
separation of the two phases. On the other hand if we cannot measure both
the volume $v$ and the energy $e$ we are left either with $P_{\beta \lambda
_{v}}\left( e\right) =\bar{W}_{\lambda _{v}}\left( e\right) Z_{\beta \lambda
_{v}}^{-1}\exp (-\beta e)$  giving access to the energy partition sum, $\bar{%
W}_{\lambda _{v}}\left( e\right) ,$ at constant $\lambda _{v}$ 
or with the probability $P_{\beta \lambda _{v}}\left( v\right) =$ $\bar{Z}%
_{\beta }\left( v\right) Z_{\beta \lambda _{v}}^{-1}\exp (-\lambda _{v}v)$
leading to the isochore canonical partition sum $\bar{Z}_{\beta }\left(
v\right) .$ Since both probability distribution $P_{\beta \lambda
_{v}}\left( e\right) $ and $P_{\beta \lambda _{v}}\left( v\right) $ are
bimodal the associated partition sum do have anomalous concavity intruders.
Both energy in the constant $\lambda _{v}$ ensemble or volume in the
canonical ensemble can be used as succedanea of the order parameter.

Let us now take another example from the Ising model. 
In the absence of a magnetic field the Ising system presents a second order
phase transition. We can now study the canonical distribution of energy $%
\hat{B}_{1}=\hat{H}$ and magnetization $\hat{B}_{2}=\hat{M}$ . 
The pertinent statistical ensemble
has two Lagrange multipliers, the canonical temperature $\alpha _{1}=\beta $
and a magnetization constraint $\alpha _{2}=$ $\beta h$ which has the
dimension of a magnetic field divided by a temperature. 
The canonical distribution of energy and
magnetization $P_{\beta }\left( e,m\right) $ is shown in Fig. 3 for three
temperatures. 
Above $T_{c}$ the distribution is normal, only the paramagnetic phase is
present. At $T_{c}$ the distribution presents a curvature anomaly on the low
energy side. Below $T_{c}$ we observe a first order phase transition, the
order parameter being the magnetization. The bimodal structure in the $m$
direction corresponds to a negative suceptibility in a constant
magnetization ensemble. It should be noticed that the projection on the
energy axis does not show anomalies. The heat capacity remains positive and
the energy cannot not be a substitute of the order parameter.


\begin{figure}[htbp]
\begin{center}
\epsfig{file=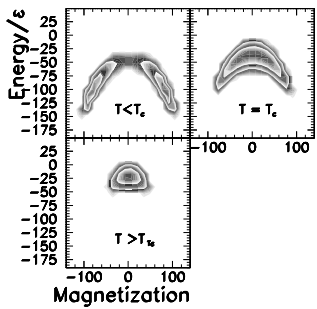,height=6.0cm}
\end{center}
\caption{Magnetization and energy distribution of an Ising model above, at
and below the critical temperature. }
\end{figure}

Finally we stress that the presented definition of phase
transition based on the probability distribution can be extended to other
ensembles of events which do not correspond to a Gibbs statistics. 
As an example, we analyze the consequence of going from Gibbs to 
Tsallis\cite{Tsallis}  
ensemble on the existence of a phase transition, 
for a system controlled by an external parameter $\lambda$ (e.g. a pressure). 
For a given $\lambda $ the system is
characterized by a density of states 
$\bar{W}_{\lambda }\left( e\right) .$ For a
critical value of $\lambda =\lambda _{c}$ the associated entropy $S_{\lambda
}\left( e\right) =\log \bar{W}_{\lambda }\left( e\right) $ presents a zero
curvature and below a convex intruder. The Tsallis probability distribution
reads ( $q_{1}=q-1)$ \cite{Tsallis}  
\begin{equation}
P_{\lambda }^{q}\left( e\right) =\frac{\bar{W}_{\lambda }\left( e\right) }{%
Z_{\lambda }^{q}}\left( 1+q_{1}\beta e\right) ^{-q/q_{1}}
\end{equation}
Computing first and second derivatives of $\log P_{\lambda }^{q}\left(
e\right) $ one can see that the maximum of $\log P_{\lambda }^{q}\left(
e\right) $ occurs for the energy which fulfills the relation $\bar{T}
_{\lambda }\left( e\right) =(\beta ^{-1}+q_{1}e)/q$ where $\bar{T}$ is the
microcanonical temperature while this point has a null curvature if
$\bar{C}_{\lambda }\left( e\right) = q/q_1$
where $\bar{C}_{\lambda}$ is the microcanonical 
heat capacity. 
Then the Tsallis critical point occurs when 
$\bar{C}_{\lambda }\left(
e\right) $ reaches $q/q_{1}$, i.e. above the microcanonical critical
point. 
Therefore, one expects a broader coexistence zone in the Tsallis ensemble
extending toward higher pressures.
Far from the critical point, the curvature at the maximum 
of $P_{\lambda }^{q}\left( e\right) $ 
is 
$\bar{T}^2 (e) 
\partial^2 \log P_{\lambda }^{q}\left( e\right) / \partial e^2 =
-1/\bar{C}_{\lambda }(e) + q_1 /q
$
Far from the $C$ divergence line, this curvature is not very different 
from the microcanonical heat capacity
since $q_1/q$ is small. 

In conclusion, we have proposed a definition of phase transitions in finite
systems based on topology anomalies of the event distribution in the space
of observations. We have shown that for statistical equilibria of Gibbs type
this generalizes all the definitions based on the curvature anomalies of
entropies or other potentials. It gives an understanding of coexistence as a
simple bimodality of the event distribution, each component being a phase.
It provides an intuitive definition of order parameters as the best variable
to separate the two maxima of the distribution or as the ridge passing
through the saddle point between the peaks associated with the two phases.
This provides an experimental tool to define the order parameter and the
existence of two phases. The nature of the order parameter provides also a
bridge toward a possible thermodynamical limit. If it is sufficiently
collective (such as one (or few) body operator) it may survive until the
infinite volume and infinite number limit. If the anomaly also survives then
the finite size phase transition may become the one known in the bulk.
Finally the proposed definition can be extended to different statistical
ensembles such as Tsallis ensemble. We have shown that phase transitions can
be identified but that the associated properties such as the position of the
critical point do change with the ensemble.

{\bf Acknowledgements}: We would like to thank all the participants and staff of
the Trento center (ECT*) workshop on "Phase transitions in finite systems"
where part of this work has been elaborated. Many thanks are due to B.
Bouriquet for the stimulating discussions on Tsallis entropy.

\end{document}